\documentclass[12pt]{iopart}
\pdfoutput=1
\usepackage{graphicx}
\usepackage{url}
\usepackage{harvard}
\usepackage[normalem]{ulem}
\usepackage{color}
\begin{document}

\title[]{Periodically bursting edge states in plane Poiseuille flow}

\author{Stefan Zammert$^1$\footnote{Corresponding 
author: Stefan.Zammert@physik.uni-marburg.de} and Bruno Eckhardt $^{1,2}$}

\address{$^1$ Fachbereich Physik, Philipps-Universi\"at Marburg, Germany}
\address{$^2$J.M. Burgerscentrum, Delft University of Technology, Mekelweg 2, 2628 CD Delft, The Netherlands}

\ead{Stefan.Zammert@physik.uni-marburg.de}

\begin{abstract}
We investigate the laminar-turbulent boundary in plane Poiseuille flow 
by the method of edge tracking. 
In short and narrow computational domains we find for a wide range  of Reynolds numbers 
that all
states in the boundary converge to a 
periodic orbit with a period of the order of $10^{3}$ time units. 
The attracting states in these small domains are periodically extended in the spanwise and
streamwise direction, but always localized to one side of the channel in the normal direction.
In wider domains the edge states are localised in the spanwise direction as well.
The periodic motion found in the small domains then induces a large variety of 
dynamical activity that is similar to one found
in the asymptotic suction boundary layer. 
\end{abstract}
\vspace{2pc}

\maketitle

\section{Introduction}
In shear flows such as plane Couette, plane Poiseuille, duct or pipe flow it is possible to observe sustained 
turbulence for Reynolds numbers where the laminar state is still stable against infinitesimal perturbations.
These flows, therefore, show a coexistence of two dynamically distinct states, a stable laminar flow, and 
a dynamically active, turbulent state. In the simplest cases each state is attractive, in which case the
boundary between them is the boundary separating the two basins of attraction. The transient 
behaviour of the chaotic states in shear flows \cite{Hof2006}
requires an extension of the concept of a basin boundary to cover transient cases as well. It was given
in \citeasnoun{Skufca2006} (see also \citeasnoun{Schneider2007} and \citeasnoun{Vollmer2009}) and uses the lifetime of perturbations as indicator.
Increasing the amplitude of the perturbation one notes a clear transition from a region in which the
lifetime varies smoothly and the trajectories return to the laminar profile, to a region with 
irregularly fluctuating lifetimes where the perturbations initially approach a state that 
has all the characteristics of a turbulent state, before eventually returning to the laminar state.
Points on the boundary between the two regions are on the ``edge of chaos", and when tracking their time evolution one
notes that they approache a state that is a relative attractor within this boundary, 
the so-called edge state \cite{Skufca2006}.

The significance of this method lies in the ease with which it provides access to exact coherent 
states. Exact coherent states play a major role in our current understanding of the transition to turbulence in 
shear flows and they can also contribute to the understanding of the turbulent dynamics \cite{Eckhardt2007b,Kawahara2012}. 
They range from the stationary states first identified in plane Couette flow 
\cite{Nagata1990,Clever1997,Waleffe1998} to the travelling waves and relative periodic orbits in pipe 
flow
\cite{Faisst2003,Wedin2004,Faisst2008,Pringle2007,Duguet2008a,Mellibovsky2012},
in duct flow \cite{Uhlmann2010} 
and in  plane Couette and Poiseuille flow  
\cite{Waleffe1998,Kawahara2001,Gibson2009,Itano2009,Gibson2013,Nagata2013a}.
Edge tracking is a robust method for identifying  coherent structures:
whereas almost all other methods require a good guess of initial conditions, edge tracking 
will invariably converge to an invariant state for just about any initial condition. 

The edge states themselves
are important for the delineation of the stable-unstable border, and for the identification of
minimal or most dangerous disturbances that trigger turbulence \cite{Cherubini2012}.
In addition, they are also indicative of the  possible behaviour of the other states around 
which turbulence forms. Specifically, it has been possible to
track a few of these states to their saddle-node bifurcation point and to follow the corresponding
upper branch states, and thereby to gain information about the states underlying the turbulent
dynamics \cite{Kreilos2012a,Avila2013}.

States on the boundary between laminar and turbulent behaviour were first identified in small
periodic domains by  \citeasnoun{Toh2003}  \citeaffixed{Toh1999,Itano2001}{see also}, followed by 
independent studies of low-dimensional models \cite{Skufca2006},
of pipe flow \cite{Schneider2007} and of plane Couette flow \cite{Schneider2008}.
For extended domains it was shown by several authors that the attracting objects in the edge of chaos are localised states \cite{Duguet2009,Schneider2010,Mellibovsky2009,Avila2013}. The connection between narrow and wide domains 
was discussed in the context of snaking bifurcations \cite{Schneider2010a} and long-wavelength instabilities
\cite{Melnikov2013}. Spatially extended flows, such as boundary layers, can also be approached with this method \cite{Cherubini2011,Duguet2012}, and states intermediate between laminar and turbulent can be identified. The spatial
growth of the boundary layer can be avoided by applying a cross flow that maintains the width of the layer,
as in the case of the asymptotic suction boundary layer (ASBL). The edge states for ASBL in narrow domains
were identified in \cite{Kreilos2013}, and further  studies revealed a rich
variety of localised but dynamically active states \cite{Khapko2013,Khapko2013a}. This dynamical activity is compatible with the possibility
that the edge states are not simple attractors but that they can be more complicated and perhaps even 
chaotic \cite{Schneider2007,Vollmer2009}.

In this paper we study plane Poiseuille flow (PPF), the pressure driven flow between parallel walls. 
To define a Reynolds number, $Re=U_{0}d/ \nu$, for this flow we use half the distance between 
the plates $d$ and the maximum velocity $U_{0}$ of 
the laminar profile in the center of the channel. 
PPF shares many features with the other shear flows, but differs in the presence of a linear instability of the laminar flow
at a Reynolds number of $Re=5772$ \cite{Orszag1971}.
The linear instability of the laminar flow creates a two-dimensional travelling wave that can be continued to lower $Re$\.
The secondary instabilities of this wave 
create 3-dimensional exact solutions studied e.g. by \citeasnoun{Ehrensteint1991}. However, these solutions are not reached with the initial conditions we use in
our edge tracking.

Below the critical Reynolds number for the linear instability, PPF shows a coexistence of transient but long 
living turbulence and a linearly stable laminar profile,
so that the technique of edge tracking can be applied and the invariant states in the edge can be identified.
For a small computational domain and at one fixed Reynolds number, \citeasnoun{Toh2003} found that the edge state is a periodic orbit. 
We will here confirm their finding and extend it to wider domains and other Reynolds numbers, where a rich
bifurcation scenario and a large variety of other edge states can be identified. 
In addition, we will show that for wide and not too long computational domains the edge states are spanwise localised orbits with intriguing spanwise dynamics.

As in other studies, we rely for our numerical simulations on the code \textit{Channelflow}, developed and maintained by \citeasnoun{J.F.Gibson2012}. 
The code uses a spectral method to simulate a doubly periodic domain with a streamwise extent $L_{x}$, a spanwise extent $L_{z}$ and a wall-normal extent of $L_{y}=2$.
It uses a decomposition of the full flow field $\tilde{\textbf{u}}$ into the laminar flow $\textbf{U}=(U(y),0,0)$ 
with $U(y)=1-y^{2}$
and the deviations $\textbf{u}$.
The flow field $\textbf{u}=(u,v,w)$ is expanded in  $N_{x}$ and $N_{z}$ Fourier modes in the streamwise and spanwise direction and in $N_{y}$ Chebychev modes in the wall-normal direction. 
In our simulation these  numbers reach up to $32 \times 65 \times 96$ in the largest domains.
The code imposes periodic boundary conditions in streamwise ($x$) and spanwise ($z$) directions and no-slip boundary conditions at the walls.  
In all calculations we impose constant mass flux. For further  details on the code we refer to the channelflow manual. 
The edge tracking algorithm is discussed in several papers, see  e.g. \citeasnoun{Itano2001}, \citeasnoun{Toh2003}, \citeasnoun{Skufca2006} or \citeasnoun{Dijkstra2013}.

The outline of the paper is as follows. In section 2, we discuss the edge states in short and narrow domains.
In section 3, we discuss the spanwise localised states and their dynamics, 
and in section 4 we end with a summary and an outlook.

\section{Edge states in short and narrow domains}
The first edge tracking calculations in PPF were carried out by \citeasnoun{Toh2003}. The domain they used was quite small and had streamwise and spanwise widths of $\pi$ and $0.4 \pi$, respectively.
Their edge trajectories are characterised by long intervals of nearly constant energy with only small variations in the flow structure and sudden bursting events that separate these intervals. 
In order to reproduce their results, we used the
same periodic domain and a spectral resolution of $N_{x}\times N_{y} \times N_{z}= 32\times 65 \times 32$.
Starting from arbitrary turbulent fields as initial conditions for edge tracking, 
we obtain edge states that show the same features as the ones
obtained by \citeasnoun{Toh2003} using especially shaped initial conditions. 

\begin{figure}
\includegraphics[]{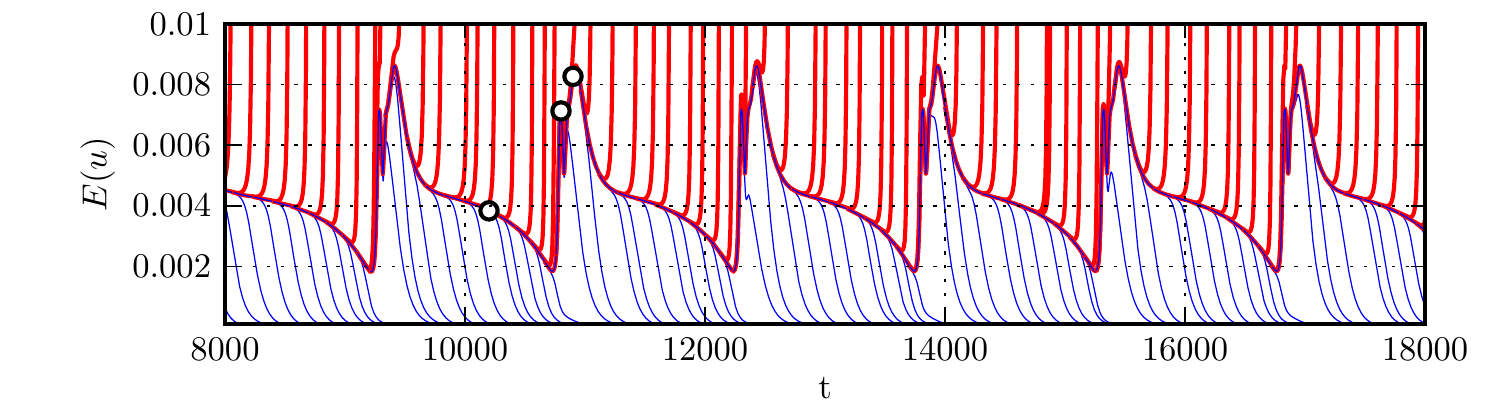}
\caption{Edge tracking at $Re=3000$ in a domain with size $L_{x}=\pi$ and $L_{z}=0.4 \pi$. For each refinement step the energies of the final turbulent and laminar trajectories are shown.
The flow in the streamwise wall-nomal plane is shown in  figure \ref{fig_YZplane_POTI} for the times marked by the black circles.
\label{fig_ET_TI}}
\end{figure}

One example for an edge tracking is shown in  figure \ref{fig_ET_TI} 
for the Toh-Itano domain size and at $Re=3000$.
The state is characterised by its energy content, 
\begin{equation}
E(\textbf{u}) = \frac{1}{L_{x}L_{y}L_{z}} \int \textbf{u}^{2} dx dy dz\,,
\end{equation} 
shown along the ordinate. The figure also illustrates how the genuine
edge trajectory is approximated by trajectories on the laminar (low energy) and on the turbulent (high energy) side 
of the edge.  As a criterion for becoming turbulent or laminar we use upper and lower thresholds for the energy.
One notes that the bursts are spaced in time by about 1000  units $(d/U_0)$. 
Given the small size of the domain and consequently the short time of 4.7 time units that a structure 
needs to traverse the domain at mean speed  $\bar{u}/U_0=2/3$, this is a very long time. 
The time on the abscissa is counted from the
start of the edge tracking, and indicates that it took several thousand 
time units to converge to this state, refered to as $PO_{1}$ in the following. We 
observed a very slow convergence to the edge state for all trajectories studied.

\begin{figure}
\includegraphics{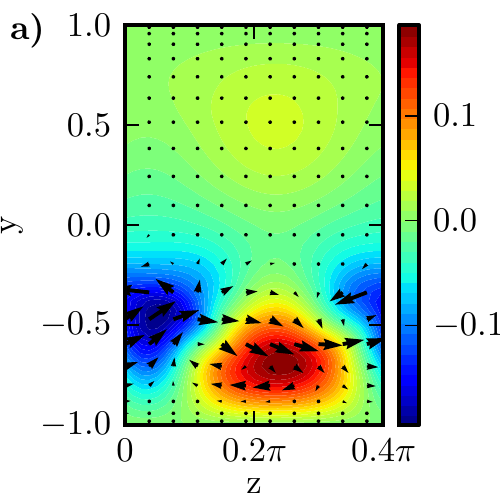}
\includegraphics{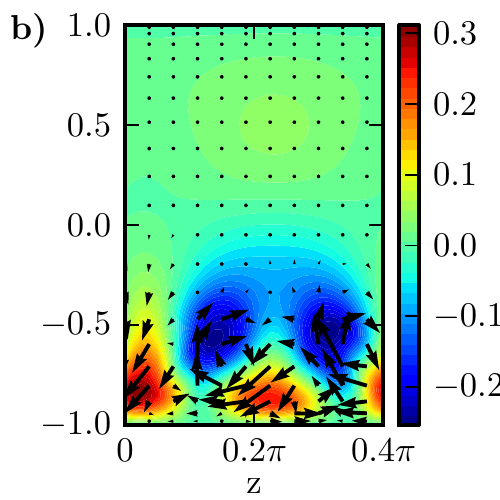}
\includegraphics{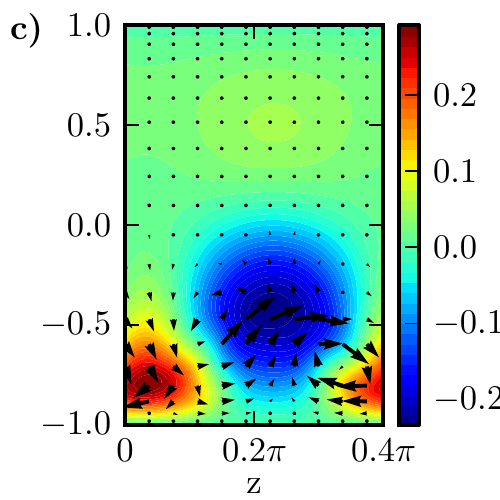}
\caption{Flow in the spanwise wall-normal  plane at $x=0$ for the times marked by the black circles in figure \ref{fig_ET_TI}. 
The streamwise velocity (deviation from the laminar profile) is colour coded (color online) and the velocity in the plane is visualised by the arrows.\label{fig_YZplane_POTI}}
\end{figure}

\begin{figure}
\includegraphics[]{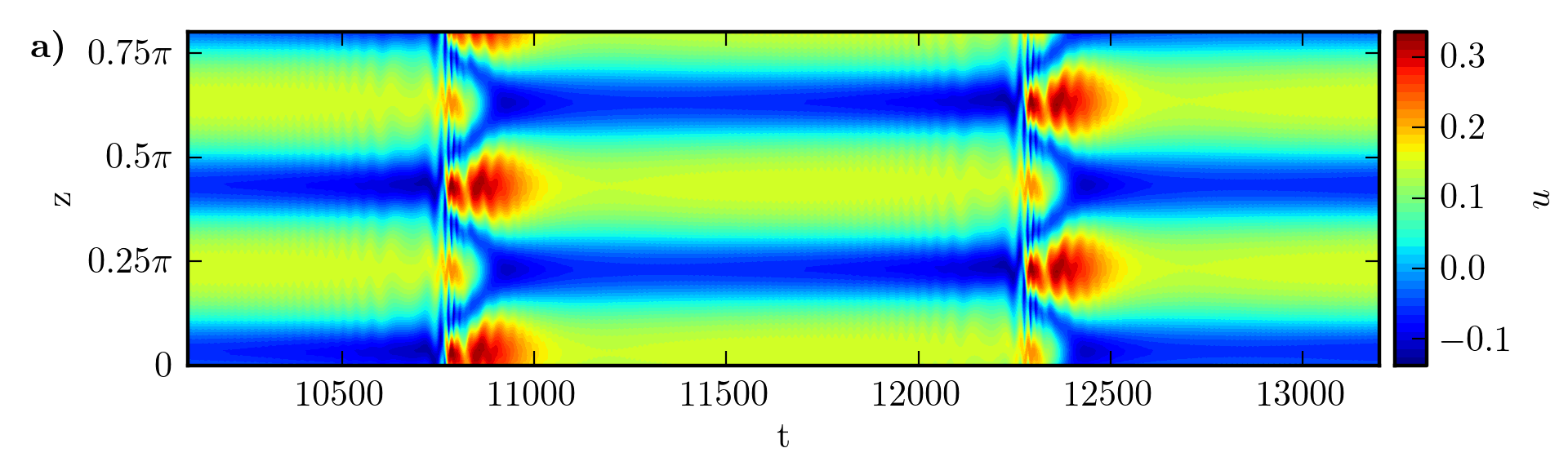}
\includegraphics[]{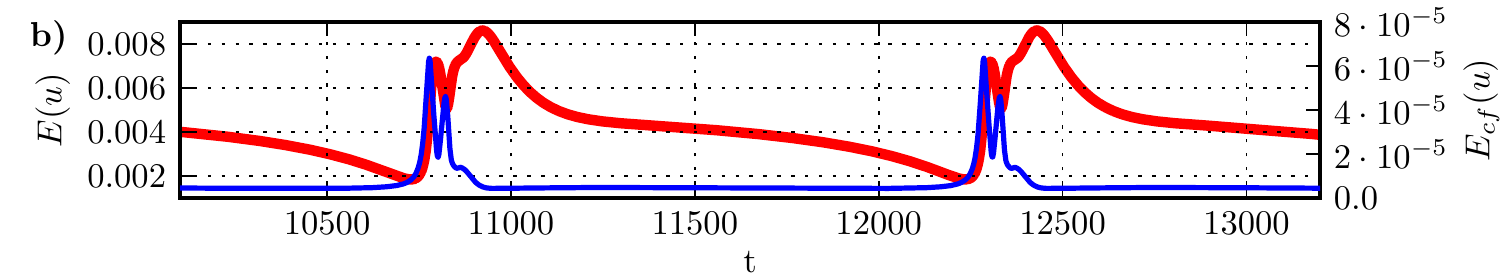}
\includegraphics[]{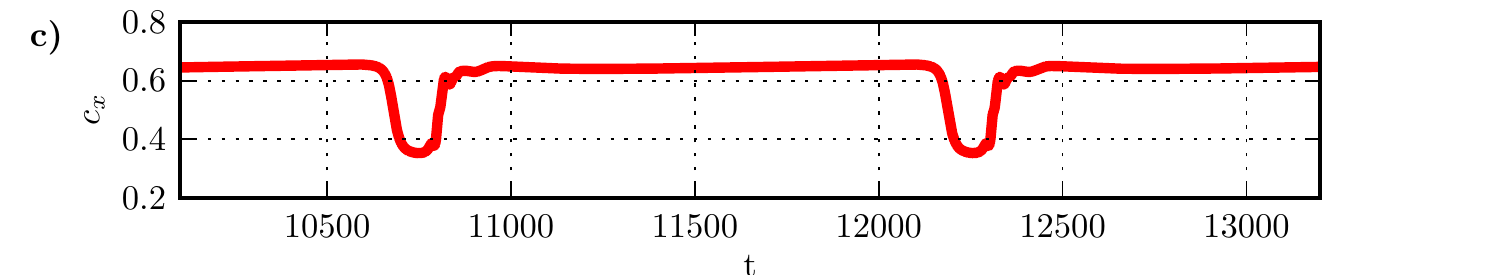}
\caption[]{Time evolution of the periodic edge state ($PO_{1}$) shown in figure 1.
Panel (a) shows the streamwise velocity at position $x=0$ in a distance of  $0.223$ from the lower wall {\it vs.} time and reveals
the spanwise dislocation by half a domain width. 
Panel (b) shows the total (thick red line) and cross-flow energy (thin blue line) for the same time interval and the massive increase
during the burst events.
Panel (c) shows the downstream advection velocity $c_x$ (determined as explained in \citeasnoun{Kreilos2013}) 
of the structures; it drops noticeably during the bursts.
\label{fig_TIstatvisualization}}
\end{figure}

Because of the regularity of the edge trajectory, \citeasnoun{Toh2003} speculate that the trajectory is attracted by a limit cycle or heteroclinic connection.
Visualisations of the flow in the spanwise wall-normal  plane at $x=0$ are given in 
figure \ref{fig_YZplane_POTI} for the times marked by the black circles in \ref{fig_ET_TI}. They show that
the state is indeed periodic, but with a period that is twice the one that one would read off from the energy vs. time diagram.
As is evident from the figure, the state consists of a pair of streaks located near one wall that changes only 
slightly over a long time interval, but then splits into two pairs of streaks, and reforms after a short 
time interval as a state with a single pair of streaks that is similar to the initial one except for a  
shift in the spanwise direction by half a domain width.
Therefore, after one period in energy one obtains the initial flow structure but shifted by half 
the domain width.
The evolution of the streak during one period is shown 
in figure \ref{fig_TIstatvisualization}(a), where the 
time-variation of the streamwise velocity at fixed $x$-coordinate and at a distance of $0.223$ 
from the lower plate is shown. 
In addition, in figure \ref{fig_TIstatvisualization}(b)  the full energy $E(\textbf{u})$ and the crossflow energy, 
$$ E_{cf}(\textbf{u}) = \frac{1}{L_{x}L_{y}L_{z}} \int (v^{2} + w^{2}) dx dy dz ,$$
are shown as a function of time.
The cross-flow energy is a more reliable indicator of persistent dynamics since the flow will relaminarize if
its value is too low. In addition, we show  
the instantaneous advection velocity $c_{x}$ of the structure as defined in \citeasnoun{Kreilos2013b}  
in figure \ref{fig_TIstatvisualization}(c). 
The instantaneous velocity varies very little over a long time interval, as is typical for trajectories
close to a travelling wave. Shortly before the bursts that displace the streaks, $c_{x}$ is reduced drastically, but 
recovers to the previous values again once the new position is reached.
The periodic orbit shows many of the features described in the 
self-sustaining process \cite{Waleffe1997}.
During the quiet phases the state consists of a pair of streaks that  decay slowly, resulting in a slow 
decrease of total energy.  At a certain point in time the streaks undergo an instability, new vortices form and the
crossflow energy content increases dramatically. The vortices then drive strong streaks that undergo 
instabilities that nucleate further streaks and the cycle starts again.

The periodic edge state is surprisingly similar to the one studied in detail by  \citeasnoun{Kreilos2013} 
in the asymptotic suction boundary layer (ASBL). The similarity of the two states can be explained by the fact 
that the base profile of the ASBL close to the wall is similar to one half of the Poiseuille profile, from the 
wall to the midplane. More specifically,  
the first two derivatives of the dimensionless profiles  of ASBL ($U(y)=1-e^{-y}$) and  PPF ($U(y)=1-y^{2}$) 
at the position of the lower wall ($y=0$ for ASBL and $y=-1$ for PPF)
are equal up to a factor of two. 
For the ASBL  the periodic edge state is created in a SNIPER 
bifurcation \cite{Kreilos2013,Tuckerman1988,Strogatz1994} 
with increasing suction velocity.
During the quiet phases the flow structure in the lower part of the channel is similar to the 
lower branch of the  well-studied NBC-solution \cite{Nagata1990,Clever1997} that is an edge state in plane Couette flow. 
The similarity becomes evident by comparing the visualisation of the NBC-state of \citeasnoun{Melnikov2013} to the lower half of the visualisation of the flow during the quiet phase shown in figure \ref{fig_YZplane_POTI}(a).

The period of the state is too long to converge it with a Newton method, so we use edge tracking to continue
the orbit in Reynolds number. A flow field of the orbit at one Reynolds number is
used as a starting point for edge tracking at a neighbouring Reynolds number. To verify that we are
still tracking the relevant edge states, we performed independent edge trackings starting from random initial conditions 
for isolated values of $Re$.
We were able to trace edge states in Reynolds numbers to about 2100.
They are similar to $PO_{1}$ in that 
all edge trajectories show bursts in energy. 
However, there are differences in detail, and for some $Re$ the sequence of bursts is not periodic 
but chaotic.

\begin{figure}
\includegraphics[]{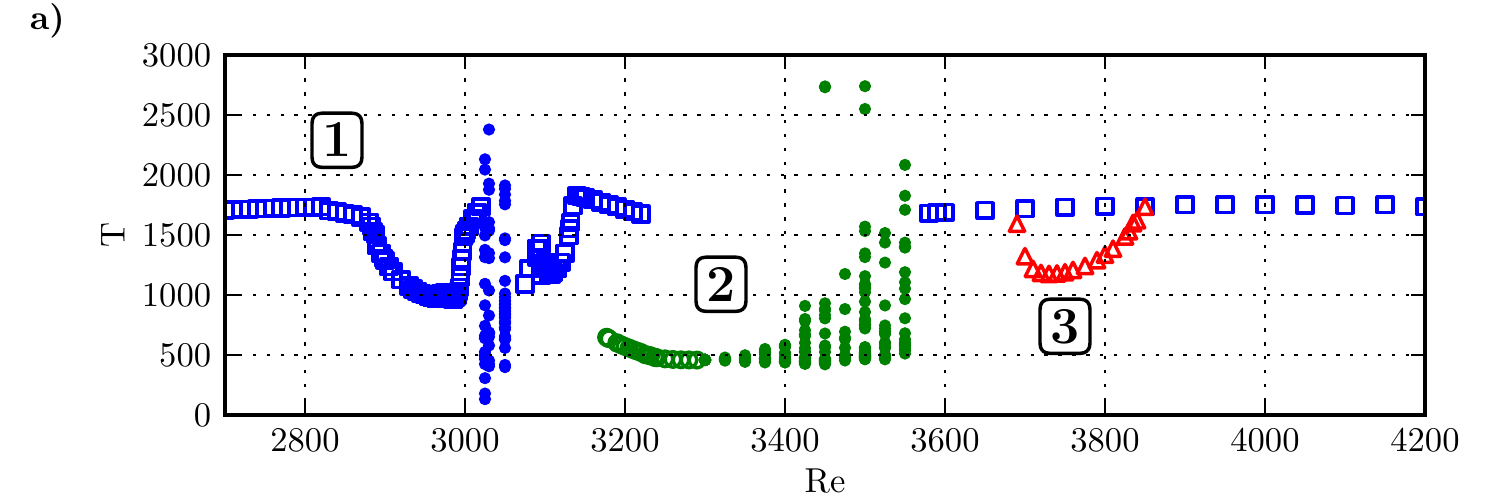}
\includegraphics[]{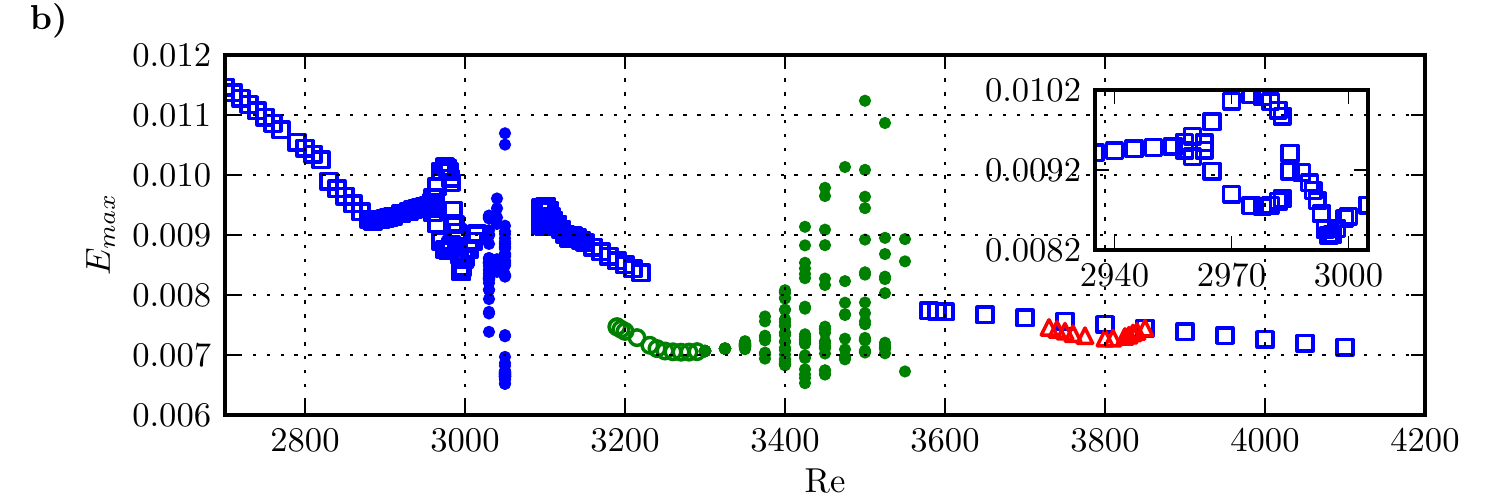}
\caption[]{Bifurcation diagram for the edge states as a function of  $Re$.
The labels $1$, $2$ and $3$ indicate where the periodic orbits $PO_1$, $PO_2$, and $PO_3$
are located.
Panel (a) shows the times between two burst. If the edge has been identified as a periodic state, the values
are shown as large symbols, if it is aperiodic small dots are shown.  
Panel (b) shows the maximum energy during the burst for the states shown in (a). The inset shows a small window
where, with increasing $Re$\, the state first undergoes a period-doubling, which then is reversed.
\label{fig_BifDiagTIstate}}
\end{figure}

In figure \ref{fig_BifDiagTIstate}(a) we show the variation of the edge state with $Re$ by 
plotting the time between two bursts along an edge trajectory.
For values of $Re$ where the attracting edge state is periodic, the distance between two bursts becomes constant
after an initial transient, resulting in a single square entry in figure \ref{fig_BifDiagTIstate}. 
For period doubled states there are two different time lapses between bursts, resulting in two squares in the
figure. For Reynolds numbers where even after a long transient time (about 20000 time steps) the time 
between two bursts varies randomly, we plot the inter-burst times using small dots;
they are then spread out over an interval of inter-burst periods.
As a second indicator we show in figure \ref{fig_BifDiagTIstate}(b) the maximum energy during the bursts.

Orbits that share with  $PO_{1}$ the property that they shift by half a wavelength in spanwise direction during one period in energy exist for various values of  $Re$. Such orbits are shown in 
 figure \ref{fig_BifDiagTIstate} using blue squares as symbols.  They undergo various bifurcations with increasing $Re$.
One example is shown in the inset in figure \ref{fig_BifDiagTIstate}(b), where a period-2 state is created in a forward bifurcation at $Re\approx 2950$ and disappears again in an inverse pitchfork bifurcation at $Re\approx2987$.

\begin{figure}
\includegraphics[]{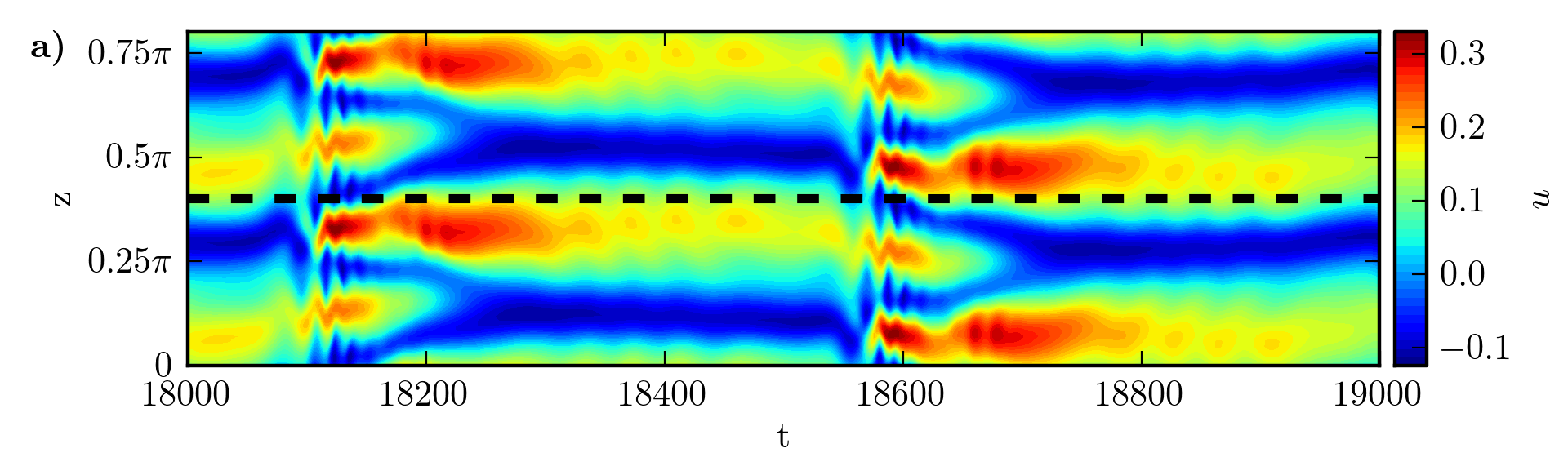}
\includegraphics[]{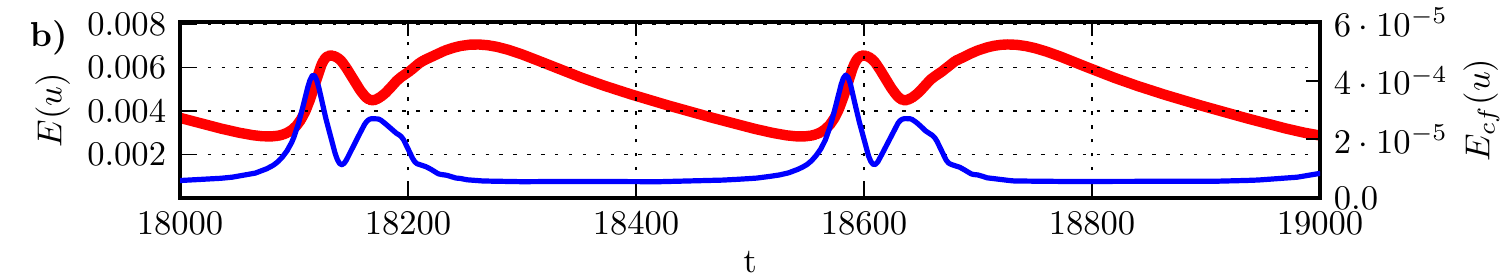}
\caption[]{Space-time dynamics for the periodic orbit $PO_2$. 
The visualisations are the same as in figure 3.
One can clearly see that the state recurs with a reflection at the dashed line. The Reynolds number is 3250. \label{fig_TIstatvisualizationR3250}}
\end{figure}

Another set of edge states, labelled $PO_2$, is found for $3200<Re<3300$ (green circles  in figure \ref{fig_BifDiagTIstate}). The states have relatively low periods between 400 and 500 time units and differ 
from $PO_{1}$ in that the structures do not shift by half a 
wavelength in spanwise direction over one period in energy. Specifically, after one period in energy the flow field equals the original one 
except for a spanwise reflection ($s_{z}: [u,v,w](x,y,z)=[u,v,-w](x,y,-z)$).
The dynamics of the $PO_{2}$-state is visualised in figure \ref{fig_TIstatvisualizationR3250}.
It undergoes a Neimark-Sacker bifurcation close to $Re=3300$, and then further bifurcations as $Re$ increases until it  finally disappears close to $Re=3600$.

\begin{figure}
\includegraphics[]{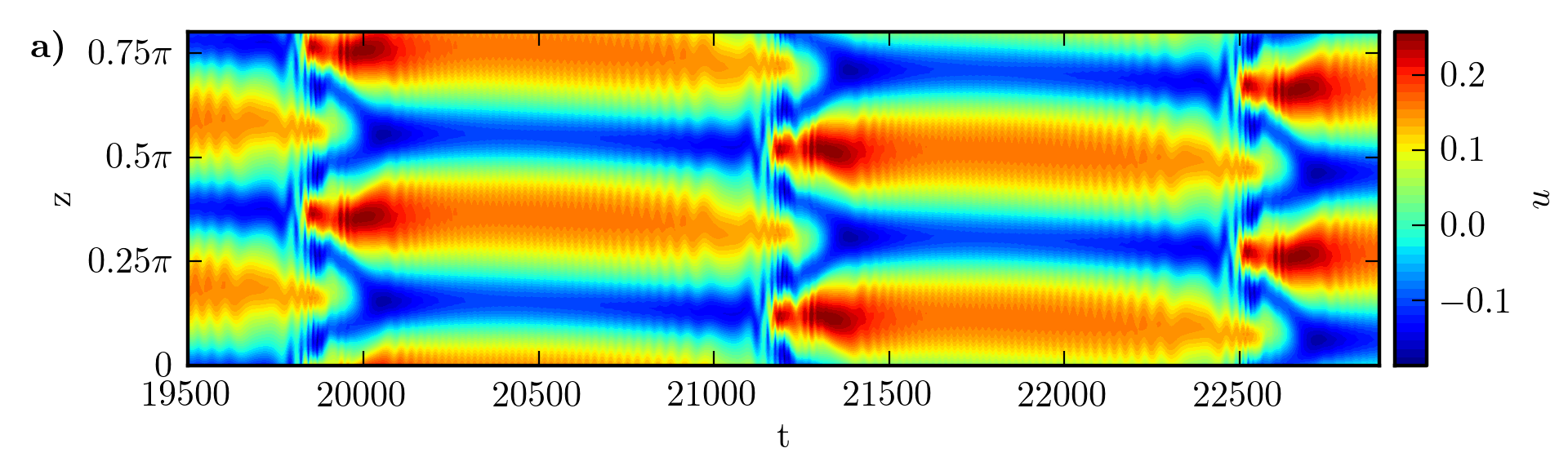}
\includegraphics[]{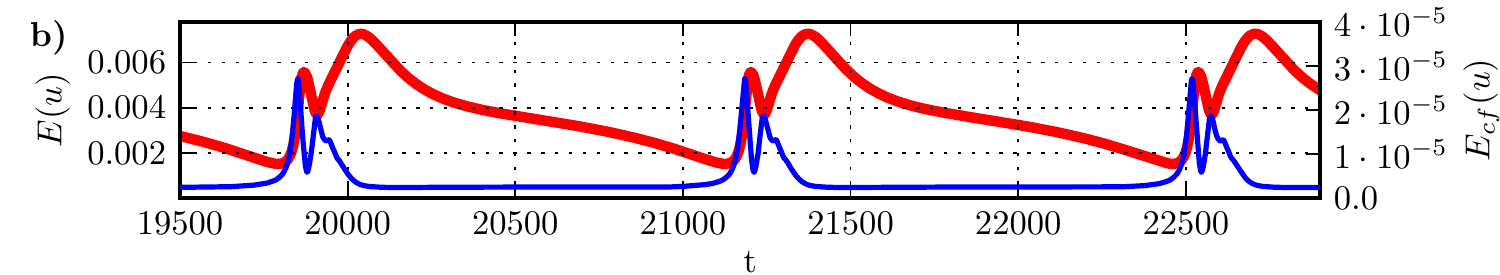}
\caption[]{Space-time dynamics for the periodic orbit $PO_3$. 
The visualisations are the same as in figure 3.
This time the displacement in the spanwise direction is $0.385L_z$.
The Reynolds number is 3800. \label{fig_TIstatvisualizationR3800}}
\end{figure}

Yet another periodic orbit, $PO_3$, is found for  $Re$ around $3800$ (red triangles in figure \ref{fig_BifDiagTIstate}). It
differs from the above orbits in the displacement following the burst, which is only $0.385 L_{z}$ 
(see figure \ref{fig_TIstatvisualizationR3800}). It seems to exist in a small range in $Re$ only. 
The deviation of the shift from half a domain width is a precursor to the behaviour in wider domains, where the
sideways displacement becomes a consequence of the intrinsic dynamics and is not correlated with the 
width of the domain (a phenomenon also seen in the ASBL by \citename{Khapko2013} \citeyear{Khapko2013,Khapko2013a}).

\begin{figure}
\centering
\includegraphics[]{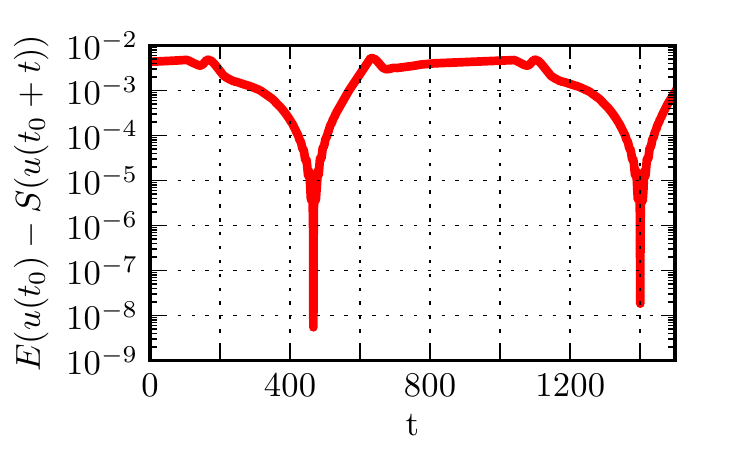}
\caption{Energy of the difference flow field $u(t_{0})-S(u(t_{0}+t))$ as a function of $t$ for $t_{0}=17964$. The symmetry operation $S$ consists 
of a reflection in spanwise direction $s_{z}$ (at the axis shown in figure \ref{fig_TIstatvisualizationR3250}) and the streamwise shift that gives the minimal difference for the value of $t$. 
The energy of the difference field has minima for $t=T$ and $t=3T$.\label{fig_ErrVsT}}
\end{figure}

In order to demonstrate the periodicity of the orbits, we consider the orbit $PO_{2}$ at $Re=3250$.  
In figure \ref{fig_ErrVsT} we show the energy of the difference between the initial velocity field
and a symmetry related one at a later time,  $u(t_{0})-S(u(t_{0}+t))$.
The symmetry operation $S$ consists of a reflection in spanwise direction 
(at the axis shown in figure \ref{fig_TIstatvisualizationR3250}) and the streamwise shift that gives the 
minimal difference for a given value of $t$.
The time between the two peaks with minimal energy is $934.05$ and corresponds to two periods $T$ in energy. Therefore,
for $Re=3250$ the period in energy of $PO_{2}$ is approximately $T=467.025$. After applying the appropriate symmetry operations the energy of
the difference of  two flow fields after a time $T$ is of order $10^{-8}$, demonstrating that the orbit is closed.
The streamwise shift for which this minimal error is achieved is $0.628 L_{x}$.
Similarly, we obtain an energy difference 
for $PO_{1}$ at $Re=3000$ of $2 \cdot 10^{-7}$ for $T=1508$ and 
for $PO_{3}$ at $Re=3800$ of $1 \cdot 10^{-7}$ for $T=1333.5$. 
The streamwise and spanwise shifts that minimize 
the error are $0.956 L_{x}$  and $0.5 L_{z}$ for
$PO_{1}$ and $0.698 L_{x}$ and $0.385 L_{z}$ for $PO_{3}$.

\begin{figure}
\includegraphics[]{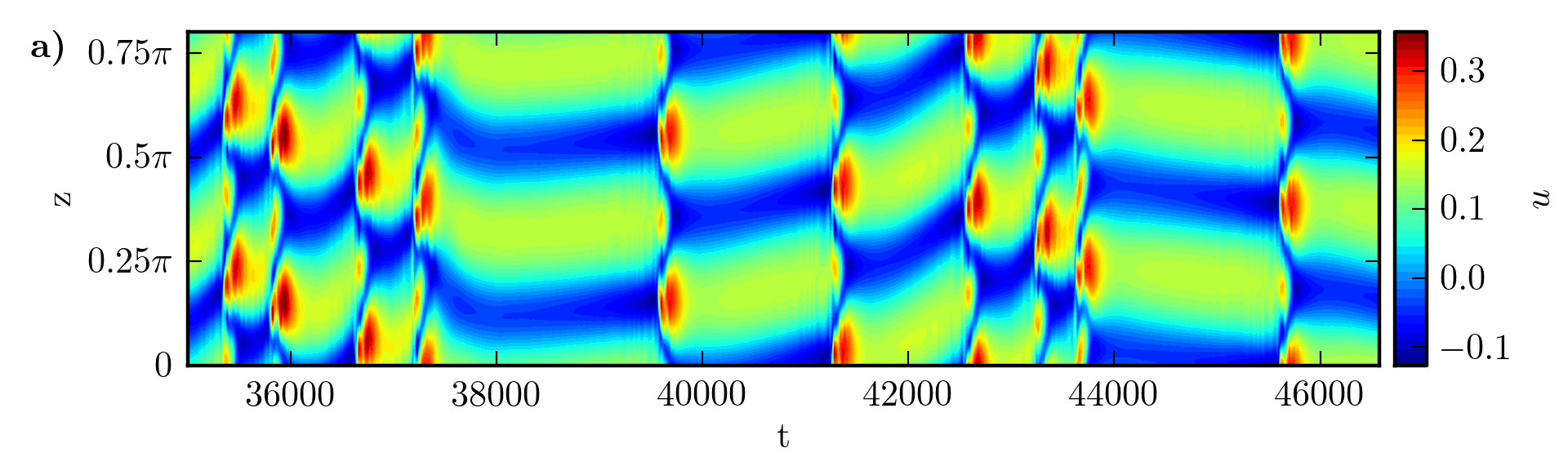}
\includegraphics[]{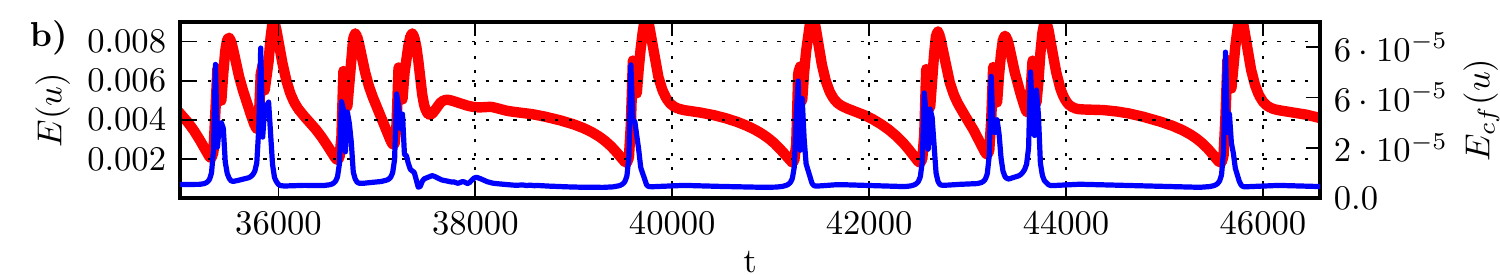}
\caption{Space-time dynamics for a chaotic edge state at $Re=3030$. 
The visualisations are the same as in figure 3.
\label{fig_TIstatvisualizationR3030}}
\end{figure}

In addition to the three orbits  $PO_{1}$, $PO_{2}$, and $PO_{3}$ highlighted above there are also 
more complicated relative attractors in the edge. In particular, in the region between $Re=3000$ and 
$Re=3100$, many different states can be found. 
Among them are period doubled states, e.g. near 
$Re=2960$ and $3095$. States with higher periods exist as well: for instance, 
at $Re=3090$ the  relative attractor shows six different maxima in energy and hence corresponds
to a period 6 state. 
For $Re=3035$ the edge trajectories are chaotic but show intermittent behaviour: the trajectories stay near a periodic state with essentially constant inter-burst period for several 1000 time units before they enter a chaotic phase.
For the slightly smaller Reynolds number of $Re=3030$ the attracting state in the edge is chaotic without any intermittency,
but the flow structure and the dynamics is quite similar to the periodic orbits at slightly higher $Re$, as 
is evident from the  visualisation of the edge trajectory shown in 
figure \ref{fig_TIstatvisualizationR3030}. 

\section{Spanwise localised edge states}
The fact that the dynamics of the plane Poiseuille edge state in 
small  periodic computational domains looks very much
like that of the edge states in the ASBL suggests a similar behaviour in wider domains. 
Specifically, we can expect that  the states become localised, and that 
the periodic sideways jumps become periodic or aperiodic translations
in the spanwise direction \cite{Khapko2013,Khapko2013a}.

\begin{figure}
\includegraphics[]{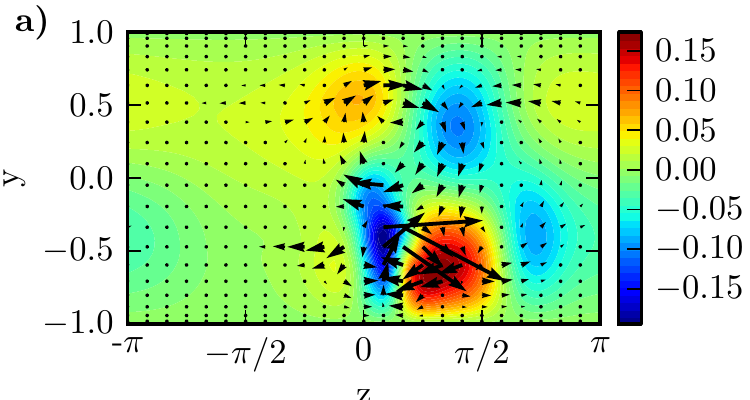}
\includegraphics[]{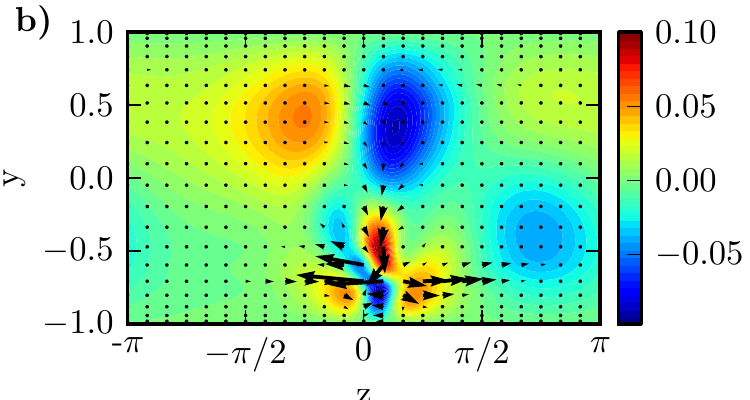}
\\
\includegraphics[]{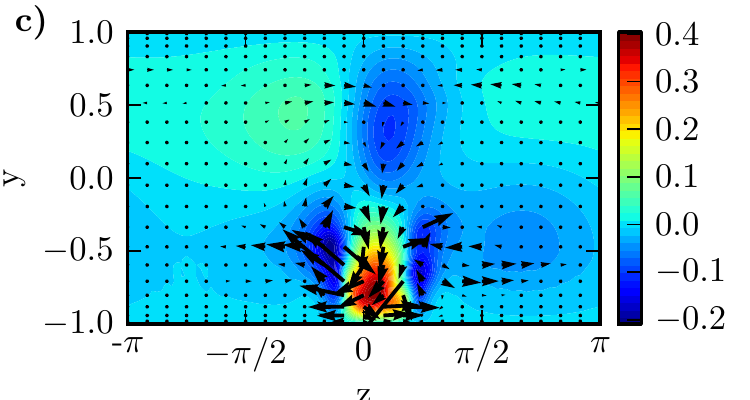}
\includegraphics[]{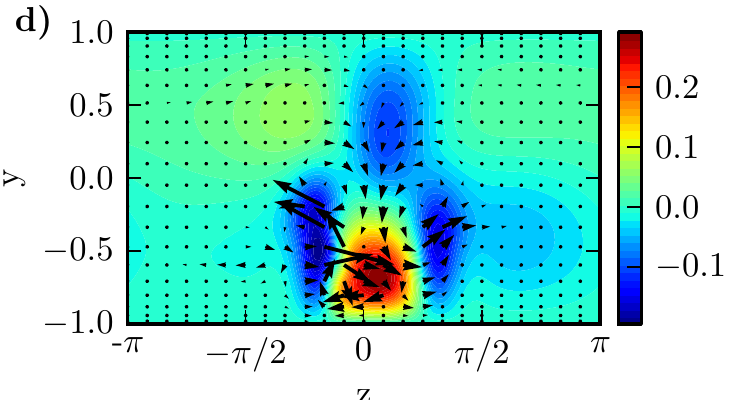}
\caption{Flow in the spanwise wall-normal plane at $x=0$ for the times marked by the black circles in figure \ref{fig_TIstatvisualization1p5pi2pi}. 
The streamwise velocity (deviation from the laminar profile) is colour coded and the velocity in the plane is indicated by the arrows.
The Reynolds number is $3000$ and the length and width of the domain are $1.5\pi$ and $2\pi$, respectively. \label{fig_YZplane_PO1p5pi2pi}}
\end{figure}

Since for PPF the domain width where the bursting periodic orbits exists is by a factor of 5 
smaller than the domain where the non-localised edge state in ASBL exists ($L_{z}=2\pi$) the width 
that is necessary for localised states might also be significantly smaller. To check this, we ran edge tracking in 
a  periodic domain with a width of $1.5\pi$ and $2\pi$ in the streamwise and spanwise direction, respectively,
and a resolution of $N_{x}\times N_{y} \times N_{z}= 32\times 65 \times 96$. The edge trajectories 
are attracted by a periodic state that is quite similar to the orbits in the small periodic domains and is in 
addition localised in the spanwise direction.
The flow visualisation  in figure \ref{fig_YZplane_PO1p5pi2pi} show that the state 
consists of a strong high-speed streak and two low-speed streaks of unequal strength to the left and 
to the right of the high speed streak.

\begin{figure}
\includegraphics{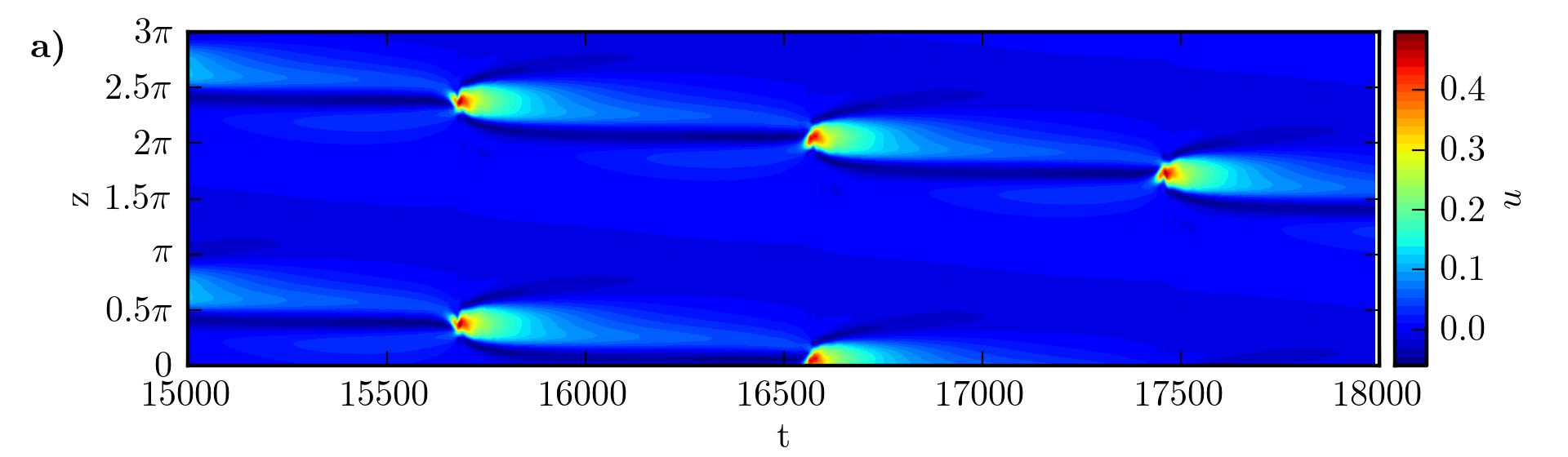}
\includegraphics{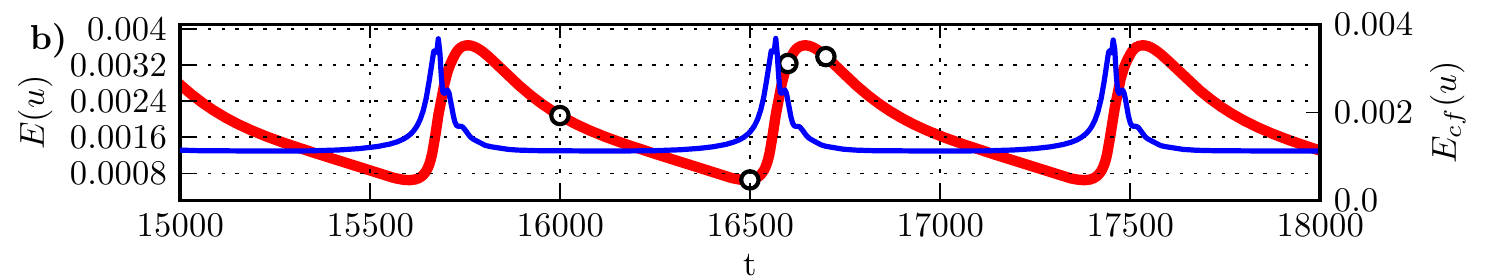}
\caption[]{Localised state in a wide computational domain.
The visualisations are the same as in figure 3, except that the velocity is recorded at $y=-0.858$.
The flow in the streamwise wall-nomal plane is shown in  figure \ref{fig_YZplane_PO1p5pi2pi} 
for the times marked by the black circles. \label{fig_TIstatvisualization1p5pi2pi}}
\end{figure}

A space-time plot of the dynamics obtained in the same manner as in the previous section is shown in figure \ref{fig_TIstatvisualization1p5pi2pi} together with the time variation of the total and the crossflow energy.
The visualisation reveals that the state shifts by approximately $0.4\pi$ in spanwise direction in each period and that the direction of the shift is always the same. In the following we will refer to this state as $PO_{E,R}$. 
By reflection symmetry, there also exists a state $PO_{E,L}$ that always  shifts in the
opposite direction. The orbits are reminiscent of the L- and R-states found 
by \citeasnoun{Khapko2013} in the asymptotic suction boundary layer (ASBL).

\begin{figure}
\includegraphics[]{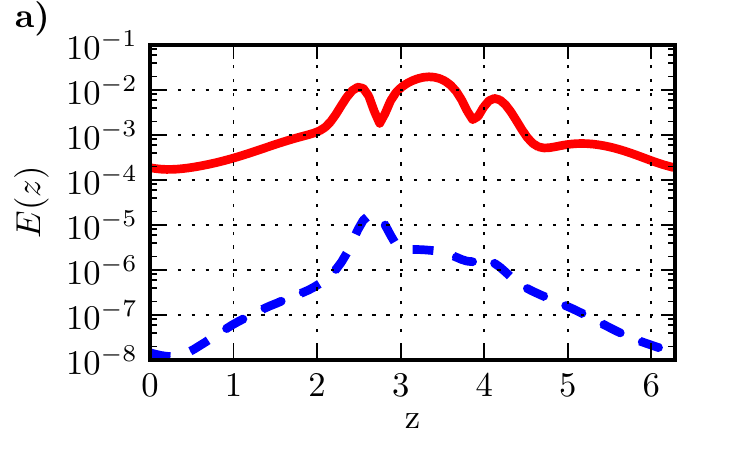}
\includegraphics[]{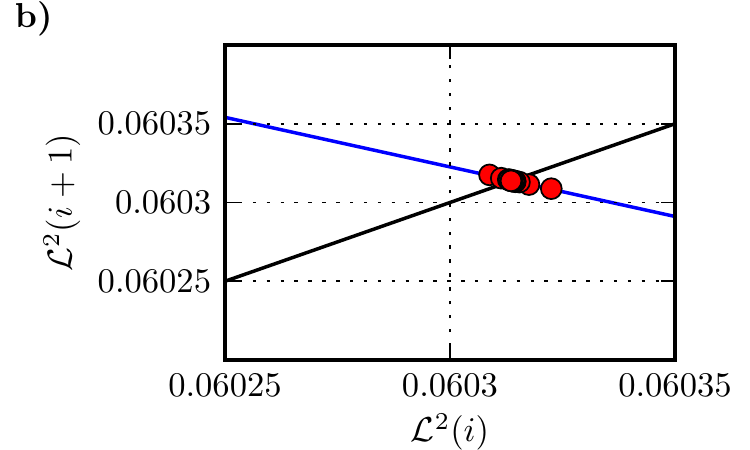}
\caption[]{Spanwise localised state $PO_{E,R}$ and its stability analysis by the method of returns.
(a) Total (solid red line) and cross flow (dashed blue line) energy density {\it vs.} spanwise coordinate $z$ for the periodic orbit at $t=16700$. 
(b) First return map of the $\mathcal{L}^{2}$-norm at its maximum value. 
\label{fig_Edensity}}
\end{figure}

A good measure that reveals the  spanwise localisation is the energy density integrated over the downstream
and normal range, but resolved along the spanwise coordinate $z$. 
The total and the cross flow energy density {\it vs.} $z$ are shown in figure \ref{fig_Edensity} for $t=16700$. 
The total energy density varies by about two orders of magnitude and the density of the crossflow energy by 
three orders of magnitude, indicating that even in this relatively narrow domain the state is already 
strongly localised in spanwise direction.

Since the periodic orbits in these domains as well as the ones in the narrower domains have very long periods,
it is not possible to track the orbits sufficiently accurately for a direct stability analysis through the
calculation of eigenvalues. Instead, we use first return maps to investigate the stability. 
We collect maxima of the $\mathcal{L}^{2}$-norm, $\mathcal{L}^{2}(\textbf{u})= \sqrt{E(\textbf{u})}$, 
and then plot the $i+1$-th maximum vs. the $i$-th maximum, as shown 
in figure \ref{fig_Edensity}(b). The slope $\beta$ of the points close to the diagonal reveals the stability. 
For $|\beta|<1$ the orbit is stable  \cite{Khapko2013a}. Indeed, for $PO_{E,R}$ the slope is about $-0.63$, so that 
we can conclude that it is a stable attractor within the edge.

\begin{figure}
\begin{center}
 \includegraphics{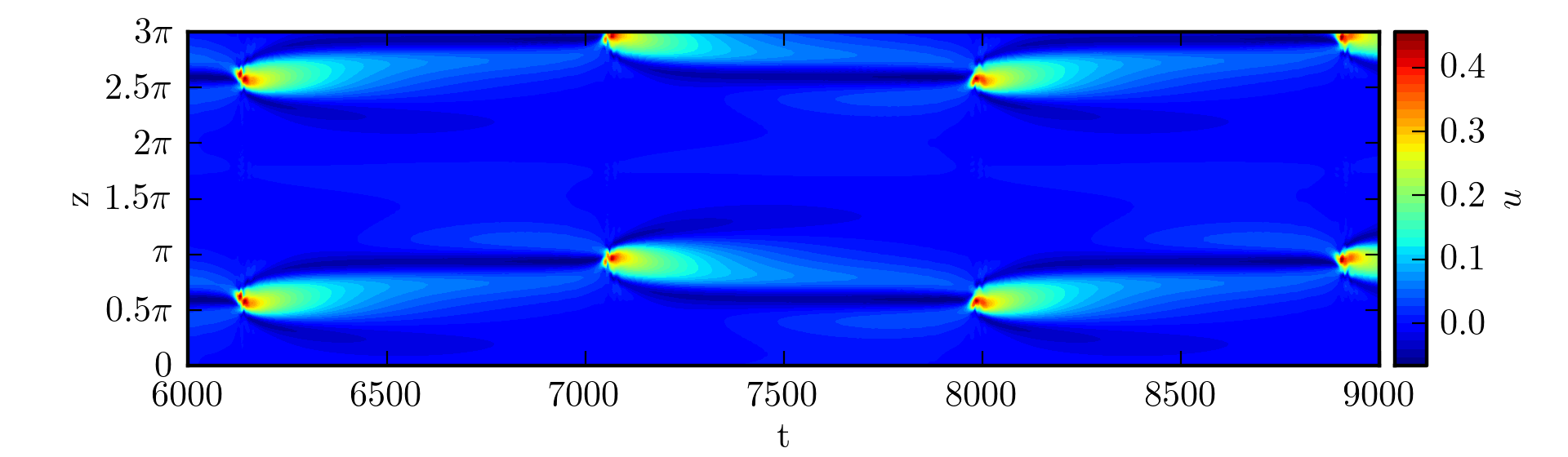}
\end{center}
\caption{Space-time representation of the periodic orbit  $PO_{E,LR}$.
Shown is the streamwise velocity at $y=-0.858$ and $x=0$ versus time.
The Reynolds number is 3000 and the length and width of the domain are $1.15\pi$ and $2\pi$, respectively.
\label{fig_TIstatvisualization1p15pi2pi}}
\end{figure}

As in the case of ASBL there are also states  with regular or irregular sequences 
of left and right displacements. 
As an example, we show in figure \ref{fig_TIstatvisualization1p15pi2pi} an
LR-state in a domain with a streamwise length of $1.15\pi$ and 
a spanwise width of $2\pi$. We will refer to this state as $PO_{E,LR}$. 
However, in contrast to the ASBL, where the R- and LR- state are both of codimension-one for 
particular values of $L_{x}$, this does not seem to be the case for $PO_{E,R}$ and $PO_{E,LR}$. 

The presence of two walls seems to have a negligible effect on the qualitative dynamics. 
The states are located near one of the walls and there are also some weaker streaks on the opposite 
side of the channel, see figure \ref{fig_YZplane_PO1p5pi2pi}, but the effect on the spanwise dynamics
is small.

When the domain length is varied the orbits $PO_{E,R}$ and $PO_{E,LR}$ undergo bifurcations that 
lead to more complex periodic states and also
chaotic ones. The continuation of the spanwise localised states in the domain length using edge tracking 
shows that $PO_{E,R}$ or $PO_{E,LR}$ or 
states that are bifurcations of these states with a more complex time dependence 
are attracting states only for $L_{x}<1.7\pi$. 
For longer domains edge trajectories show periodic behaviour that is similar to the 
one of $PO_{E,R}$ or $PO_{E,LR}$ only transiently.

\section{Conclusions and Outlook}
The study of edge states in plane Poiseuille flow has revealed a rather rich variety of states. Some of the findings
could be expected: the presence of a maximum in the flow speed in the centre separates two regions of 
opposite shear and helps states to be localised close to one side of the channel. Then their dynamics is 
similar to that in other shear flows, and in particular similar to the one in the ASBL. The fact that 
the surprising richness in spanwise localised states that show complex sequences of discrete jumps 
in the spanwise position is found not only in ASBL but also here in plane Poiseuille flow suggests
that there should be a robust mechanism for this dynamics that does not depend on the specific details
of the flow. Perhaps the one feature that could be important is the combination between a rigid wall
and a softer upper boundary where the shear vanishes: this is the case at the midplane in plane Poiseuille flow
and at the upper end of the boundary layer in the  ASBL. 

The similarity between the dynamics found here in the plane Poiseuille flow and the previous results for the ASBL 
raises the question to which extend they can be connected. For this one can look for homotopies that transform
one flow into the other. \citeasnoun{Kreilos2013} show how ASBL and plane Couette can be connected and how
the periodic solution appears in a SNIPER-bifurcation of the continuation of a simple equilibrium solution on the 
way from PCF to ASBL. 
\citeasnoun{Waleffe2003} uses a homotopy between PPF and PCF to relate travelling waves and steady states. 
So one can expect that a combination of both homotopies can provide a connection between the ASBL and plane Poiseuille flow, but we did not pursue this further.

Several directions for further investigation are evident. It would be most interesting to continue some of these
states to the saddle-node bifurcation points so that the upper branch can be identified and one can see
how they impact the scaffold for turbulent dynamics (along the lines discussed by \citeasnoun{Kreilos2012a} and \citeasnoun{Avila2013}). 
Similarly, it would be interesting to track them with a Newton method and to trace the orbits 
without the need for edge tracking, so that one can follow the states that branch of at bifurcations.
That, however, is challenging  and computationally expensive 
because of the long periods of the orbits which require multi-point shooting methods \cite{Sanchez2010}.

\section*{Acknowledgements}
We thank Tobias Kreilos for discussions, and John Gibson for providing {\it Channelflow}.
This work was supported by the Deutsche Forschungsgemeinschaft.

\section*{References}


\bibliographystyle{jphysicsB}
\end{document}